\documentclass{article}
\usepackage{spconf,amsmath,graphicx}
\usepackage{subcaption}
\usepackage{spconf,amsmath,graphicx,amssymb,bm,textcomp,booktabs}
\usepackage{url}
\usepackage{subcaption}
\usepackage{siunitx}
\usepackage{cite}
\usepackage{caption}
\usepackage{tablefootnote}
\usepackage{tipa}
\title{An Empirical Study on L2 Accents of Cross-lingual Text-to-Speech Systems via Vowel Space}
\name{Jihwan Lee, Jae-Sung Bae, Seongkyu Mun, Heejin Choi, Joun Yeop Lee, Hoon-Young Cho, Chanwoo Kim}
\address{Samsung Research, Seoul, South Korea}
\begin{document}
%\ninept
%
\maketitle
\begin{abstract}

With the recent developments in cross-lingual Text-to-Speech (TTS) systems, L2 (second-language, or foreign) accent problems arise.
Moreover, running a subjective evaluation for such cross-lingual TTS systems is troublesome.
The vowel space analysis, which is often utilized to explore various aspects of language including L2 accents, is a great alternative analysis tool.
% Also, running a subjective evaluation for such cross-lingual systems is troublesome. 
% Abeysinghe, et. al. in 2022 propose to utilize the vowel space analysis as an intermediate evaluation tool to visualize accents of English.
In this study, we apply the vowel space analysis method to explore L2 accents of cross-lingual TTS systems.
Through the vowel space analysis, we observe the three followings: a) a parallel architecture (Glow-TTS) is less L2-accented than an auto-regressive one (Tacotron); b) L2 accents are more dominant in non-shared vowels in a language pair; and c) L2 accents of cross-lingual TTS systems share some phenomena with those of human L2 learners.
Our findings imply that it is necessary for TTS systems to handle each language pair differently, depending on their linguistic characteristics such as non-shared vowels. They also hint that we can further incorporate linguistics knowledge in developing cross-lingual TTS systems.

\end{abstract}
\section{Introduction}
\label{sec:intro}

Over the recent years, various approaches have been proposed to build cross-lingual Text-to-Speech (TTS) systems \cite{Zhang2019, nachmani2019unsupervised, ye2022improving}.
% However, two major issues arise in developing such cross-lingual TTS models.
However, L2 (second-language) accents frequently appear in such cross-lingual scenarios, and several attempts have been made to improve its nativeness \cite{Zhang2019, nachmani2019unsupervised, ye2022improving}.
% L2 accents, also known as foreign accents, in multi-lingual tts have been reported \cite{}.
% And many attempts to mitigate this L2 accent problem in multi-lingual tts. \cite{}
% For instance, 
To make matters even worse, subjective evaluation of less major languages is challenging, especially for researchers in a less diverse environment.
Running a subjective evaluation test every time can be a hassle, unless you are a fluent polygot.

% SKMUN 아무말-Start
% < VER 1>
% (1) Multi-lin TTS 많이함
% (2) 근데 엑센트 어색 이슈 있 [1-3]
% (3) 이 이슈를 어찌저찌 풀려함 [1-3]
% (4) 근데 타임/디비/effort consuming
% (5) 우리가 nice-alternative 어프로치 로 L2 제안함

% < VER 2>
% (1) 음운(음소?)론에 vowel space 분석이라는게 있음
% (2) 뉴질랜드가 intra-language (dialect) 분석용으로 씀
% (3) 이 연구를 확장하여 inter-lang 도전
% (4) 이를 통해 우리는 기존 multi-lin TTS가 고질적으로 갖는 엑센트 어색 이슈를 "분석" 할수 있 [1-3]
% (5) More specific, 우리는 아래의 퀘스쳔에 집중하여 리서치를 해보려 함 
% 1. 2. 3.

% 이렇게 쓰면 솔직하게 (?) 우리가 빌드업 해온 흐름은 잘 사는데
% 단점은 뉴질랜드 논문 v2 느낌이 나서.. 뭔가 지협적인 문제를 푼다는 뉘앙스가 있음

% SKMUN 아무말-End

The vowel space analysis method can be a nice alternative.
The vowel space refers to the two-dimensional area depicting the jaw opening and the tongue positions of vowels \cite{fant1973speech}.
It is accompanied by the formant analysis, since the jaw opening and the horizontal position of the tongue are highly related to the first and second formant frequencies ($F1$ and $F2$), respectively \cite{fant1973speech, sandoval2013automatic}.
It is utilized to explore various aspects of language, including L2 accents \cite{dowd1998learning,kartushina2014effects, kartushina2016mutual, huffman2020relation, nycz2013best,dimov2013non}.
% The vowel space can even function as a visual aid to new language learners, or L2 learners, by providing a pronunciation guide to the correct tongue positions \cite{dowd1998learning}.
% The vowel space can even provide a useful visual guide to new language learners, by correcting the tongue position \cite{dowd1998learning}.

Abeysinghe, et al. \cite{abeysinghe22_interspeech} propose to utilize the vowel space analysis as an intermediary evaluation tool to assess accents or dialects of English.
% Abeysinghe, et. al. \cite{abeysinghe22_interspeech} propose to use the vowel space as an intermediary evaluation tool to assess accents or dialects of English.
% The vowel space is a two-dimensional diagram 
% They utilize the vowel space analysis to show how 
% As Abeysinghe \cite{abeysinghe22_interspeech} points out, linguists utilize the vowel space to explore various aspects of language, including L2 accents \cite{}.
% The vowel space can also function as a visual aid to new language learners, or L2 learners, by providing a pronunciation guide to the correct tongue positions.
In this paper, we extend their vowel space analysis approach \cite{abeysinghe22_interspeech} to cross-lingual TTS systems.
We propose a cross-lingual vowel space analysis method and probe L2 accents in cross-lingual TTS systems.
To our best knowledge, this is the first attempt to utilize the vowel space analysis to investigate the L2 accents of cross-lingual TTS systems.

% Utilizing the vowel space analysis, we show that the L2 accents are more dominant in the not shared vowels than the shared ones.

In this work, we focus on the following research questions, by utilizing the vowel space analysis in cross-lingual scenarios:
\begin{enumerate}
    \item Do L2 accents of cross-lingual TTS systems differ by model architectures?
    \item How are L2 accents of cross-lingual TTS systems different by linguistic characteristics among languages?
    \item Is any of L2 accent observations in cross-lingual TTS systems also present in linguistics literature conducted on actual human L2 learners?
    % Are L2 accent characteristics of TTS models similar to those of actual human L2 learners?
    % \item 
\end{enumerate}
To answer the questions above, we choose to explore cross-lingual TTS systems based on Tacotron \cite{taco1, taco2, ellinas2020high} and Glow-TTS \cite{kim2020glow} as representative backbone models of auto-regressive and parallel (non auto-regressive) architectures, respectively.
We assess the following two factors that are commonly explored in the vowel space analysis \cite{kartushina2014effects, kartushina2016mutual, huffman2020relation, nycz2013best, dimov2013non}: the accuracy and compactness of vowel categories.
It has been reported that these two factors are highly correlated with L2 accents \cite{huffman2020relation, kartushina2016mutual}.
The vowel accuracy is the correctness of the position of a synthesized vowel, compared to its native one on the vowel space.
% We calculate the vowel accuracy by measuring a Euclidean distance between the median points of a non-accented (native) vowel and its cross-lingual version on the vowel space.
The vowel compactness, or the vowel variability, is one way to assess acoustic stability among realizations of each vowel \cite{kartushina2016mutual}.
% We evaluate the vowel compactness by computing the standard deviation of each vowel on the vowel space.
% Various linguistics papers imply that the greater vowel distance and 
% We apply different speaker/language embedding configurations to each model.
% Our observations indicate that Glow-TTS is less prone to L2 accents than Tacotron in general.
% and adding the language embedding to the model proves its effective in Tacotron, but no clear evidence is shown in Glow-TTS.
% To answer the second question, we conduct experiments based a database containing multiple languages.
% : German (DE), French (FR), English (EN), Spanish (ES), Korean (KR), and Japanese (JA).
% We also compare if non shared vowels are more vulnerable.

Evaluating these metrics, we compare the shared and non-shared vowels in a language pair.
The shared vowels refer to the vowels that exist in both languages of a language pair, while the non-shared vowels do not.
We inspect some of the vulnerable cases in cross-lingual synthesis, regarding these linguistic characteristics.
% probe these linguistic characteristics and inspect some vulnerable cases in cross-lingual synthesis.
% Our experimental results also show that L2 accents are more dominant in the non-shared vowels in a language pair.
% Furthermore, L2 accents are most severe when the target language is German, and when the source language is Japanese or Spanish.
% In addition to that, we survey if any of our observations in the cross-lingual TTS systems appears in linguistics literature as well.
In addition, we investigate some of the L2 accent characteristics from TTS systems, that are also present in experimental results conducted on actual human L2 learners.

\section{Methodology}
\label{sec:methodology}

\subsection{Database}
We utilize part of the CSS10 datasets \cite{park2019css10} along with the LJ Speech dataset \cite{ljspeech17} and our in-house datasets recorded by professional voice actors. 
The total database consists of three American English (EN) speakers, two Korean (KO) speakers, and one speaker from each of the following languages: German (DE), Spanish (ES), French (FR), and Japanese (JA).
Table \ref{tab:database} shows the detailed database information.
% is shown in Table \ref{tab:database}.
% They consist of total 33333 utterances with 333 hours of recordings.
We randomly select 300 utterances from each speaker as the test set.

\subsection{TTS systems}

We investigate two mainstream architectures of TTS systems: auto-regressive and parallel (non auto-regressive).
We choose a Tacotron \cite{taco1, taco2} variant \cite{ellinas2020high} as the representative model for the former, and Glow-TTS \cite{kim2020glow} for the latter.
They are suitable for cross-lingual analysis, as no external forced-alignment is required. 
% Both require no external duration information.
% We utilize TTS models based on two different architectures: a) a Tacotron \cite{taco1, taco2} variant \cite{ellinas2020high}; and b) Glow-TTS \cite{kim2020glow}.
% We choose these two, because Tacotron is auto-regressive, while Glow-TTS is parallel (non auto-regressive). 
The Tacotron variant is a slightly modified version, utilizing the same acoustic features as the LPCNet \cite{valin2019lpcnet}.
Glow-TTS is also slightly adjusted to take the same acoustic features as input.
In this paper, for simplicity, we refer this Tacotron variant and the modified Glow-TTS as Tacotron and Glow-TTS, respectively.
% We use the same acoustic features to Glow-TTS in the experiments.
We use the bunched LPCNet2 \cite{park22_interspeech} as a neural vocoder to transform acoustic features into actual waveforms.

We train these systems in single speaker and multi-lingual versions.
% ; and c) multi-lingual.
In the single speaker version, we use the base architecture from each model \cite{ellinas2020high, kim2020glow}, without any speaker or language embedding, only one speaker's data being separately trained.
This single speaker version is essentially mono-lingual, therefore we assume that no L2 accent is present in this condition.
Hence, we set this single speaker version as the non L2-accented anchor.
% In the multi-speaker condition, we condition a speaker embedding from a look-up table, utilizing all of utterances from all speakers and languages in the train set together.
The multi-lingual version is an extended version of the single speaker version.
In the multi-lingual version, we utilize the speaker embedding and the language embedding from a look-up table.
In the case of Tacotron, the speaker and language embeddings are fed to the decoder as in the \cite{Zhang2019}.
In Glow-TTS, the speaker embedding is applied as in the original architecture \cite{kim2020glow}, and the language embedding is concatenated to the speaker embedding.
% when it is being utilized.
% We assume that utilizing the language embedding along with the speaker embedding improves its naturalness than when only the speaker embedding is solely utilized, as proposed by many previous researches \cite{}.
We train the Tacotron single speaker version for 350k steps, and the multi-lingual version for 500k steps.
Glow-TTS versions usually require more training steps, so we train the Glow-TTS single speaker version for 500k steps, and the multi-lingual version for 1000k steps.
% We utilize one NVIDIA A100 GPU for each model configuration.
Other training configurations are identical to the original architectures \cite{ellinas2020high, Zhang2019, kim2020glow}.

We utilize the International Phonetic Alphabet (IPA) \cite{international1999handbook} as phoneme input to the TTS systems.
% modelfor phonemes as input text to the models.
In the multi-lingual versions, we add a language tagging to each phoneme to differentiate the same IPA phoneme in different languages, as one IPA phoneme's actual acoustic realization as a phone may differ in each language.
For instance, the phoneme input /i/ in English and the same phoneme input /i/ in Korean are considered different input tokens.
We limit our inspection scope to monophthongs, since analyzing diphthong requires a time-dynamic analysis.

\begin{table}[t]
    \small
  \caption{The number of utterances, the duration, and the number of vowels (monophthongs)\protect\footnotemark
%   {The exact number of vowels may vary by region and reference.
%   Here, we list the number of vowels (monophthongs) in our experimental setting.
%   } 
  \cite{kleiner2015duden, collins2013practical,ladefoged2014course, kwak2003vowel, vance2008sounds} of each language in the database.}
  \label{tab:database}
  \centering
  \begin{tabular}{c  c c c c}
    \toprule
    \multicolumn{1}{c}{\textbf{Language}} & 
                                         \multicolumn{1}{c}{\textbf{Code}} & \multicolumn{1}{c}{\textbf{Utterances}} & \multicolumn{1}{c}{\textbf{Duration (h)}}&\multicolumn{1}{c}{\textbf{\# of Vowels}}\\%&

        \midrule

    German&DE&7,427&16.7&17\\
    English&EN&39,455&59.5&12\\
    Spanish&ES&11,110&23.8&5\\
    French&FR&8,649&19.2&14\\
    Japanese&JA&6,841&14.9&5\\
    Korean&KO&24,962&34.0&7\\

    \bottomrule
  \end{tabular}
  
\end{table}
\footnotetext{The exact number of vowels may vary by region and reference.}

\subsection{Vowel space analysis method}
We extend the previous vowel space analysis approach \cite{abeysinghe22_interspeech} from mono-lingual to cross-lingual scenarios.
% To our best knowledge, this is the first attempt to utilize the vowel space analysis to examine the L2 accents of cross-lingual TTS models.
Our vowel analysis method comprises the following steps.
First, we extract the first and second formants of synthesized vowels.
Next, we apply speaker normalization to the formant values.
Lastly, we measure vowel accuracy and compactness for analysis.

\subsubsection{Formant estimation}

We first begin with extracting formant values.
To synthesize vowels from the TTS systems, we give text input as a sequence of tokens, beginning with a target vowel followed by a silence token and a subsequent random sentence starting with a voiceless plosive consonant such as /p/, /t/, or /k/.
Afterwards, we slice the synthesized speech by the first silence boundary and take the first segment.
We set the threshold energy as 10 dB.
We empirically observed that appending a sentence beginning with a voiceless plosive consonant induces a clearer silence boundary, easing the automatic slicing process.
As in \cite{abeysinghe22_interspeech}, the $F1$ and $F2$ values are extracted at the middle point of the chosen segment.
% using the \texttt{PRAAT}\footnote{https://www.fon.hum.uva.nl/praat} default configuration.
We synthesize 100 samples per vowel, and set the median as the representative.

\begin{figure}[t]
    \centering
    \includegraphics[width=\linewidth]{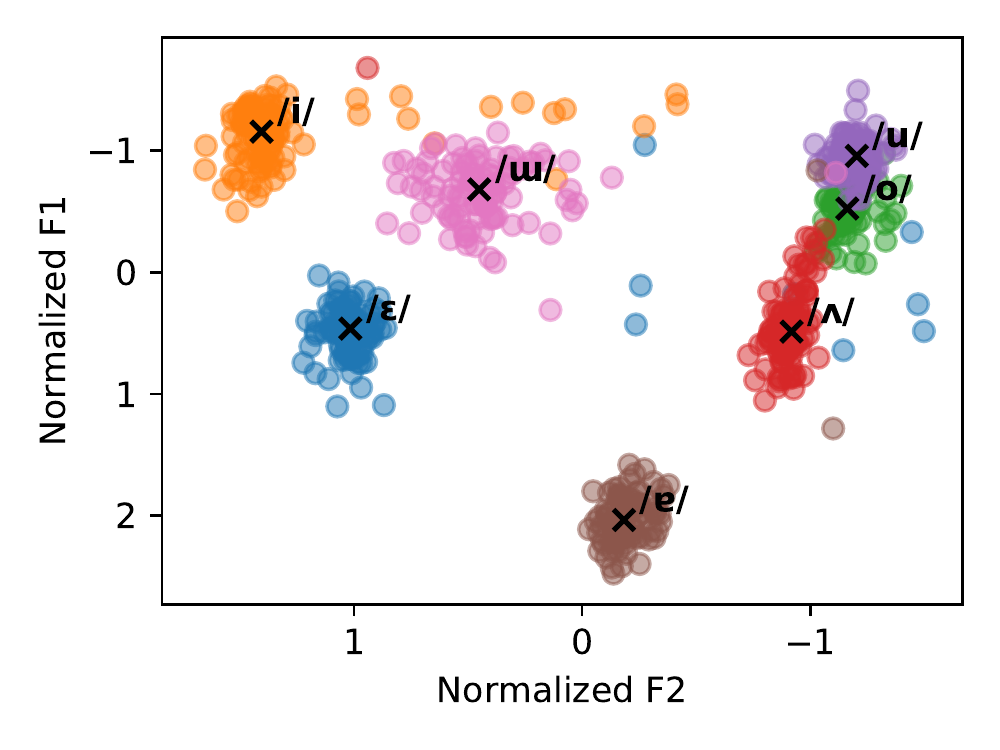}
    \caption{The normalized vowel space of a male Korean speaker which corresponds to the Korean phonology \cite{kwak2003vowel}.}
    \label{fig:vowel_space_kr}
\end{figure}

% We place the vowel in the beginning to minimize any error propagation from previous text tokens in the auto-regressive case, so we can make a fair comparison between the auto-regressive and non auto-regressive models.

Our approach has some advantages over the previous method \cite{abeysinghe22_interspeech}.
% We do this way due to the following reasons.
First, it requires no external aligner or segmentation, thus it is free from any potential errors caused by an external segmentation tool. 
It can also be utilized to low resource languages where no aligner or segmentation tool is publicly available.
Second, it minimizes the co-articulation effect by nearby phonemes in the cross-lingual scenarios.
For example, the conventional hVd test \cite{hillenbrand1995acoustic} does not apply equally to some languages such as French and Japanese, where syllable-initial /h/ and syllable-final /d/ are not allowed, respectively. 
Third, it is closer to the actual usage scenario of TTS systems than the previous method.
In the previous approach \cite{abeysinghe22_interspeech}, multiple hVd words with an in-between pause are given as text input. However, this is slightly astray from what a TTS system usually is trained for.
Compared to this, our method is more coherent with the real world usage of TTS systems.

\subsubsection{Speaker normalization}

The raw numerical $F1$ and $F2$ values in the vowel space vary by speaker \cite{lobanov1971classification}.
In order to extend Abeysinghe et al.'s approach \cite{abeysinghe22_interspeech} to cross-lingual TTS systems, a speaker normalization process should precede.
We apply the renowned Lobanov normalization method \cite{lobanov1971classification} to normalize speaker information on the vowel space, as many other linguistics researchers have proven its effectiveness \cite{adank2004comparison, fabricius2009comparison}.
% The Lobanov normalization is as follows:
% \begin{equation}
%     F_{i}^{norm} = \frac{F_i - \mu_i}{\sigma_i}
% \end{equation}
% where $F_i$ is the $i$\textsuperscript{th} formant; $\mu_i$ and $\sigma_i$ are the mean and the standard deviation of $F_i$, respectively.

\subsubsection{Vowel space visualization}
After the speaker normalization process, we visualize the vowels on the vowel space diagram, where the normalized $F2$ and $F1$ values are depicted in the reversed $x$ and $y$ axis, respectively.
Figure \ref{fig:vowel_space_kr} shows the vowel space of a male Korean speaker after the formant estimation and speaker normalization process. The positions of the vowels in the vowel space correspond to those from Korean phonology \cite{kwak2003vowel}.

\subsubsection{Vowel accuracy and compactness}

% \subsubsection{Probe \#1: vowel correctness}

% We first explore the vowel distances 

% \subsubsection{Probe \#2: vowel compactness}

% The vowel compactness 
% It has been repeated reported that L2 accent perception is highly correlated with the compactness of vowel categories \cite{}.
% We measure mean standard deviation of 

We focus on the following two factors that are commonly investigated in the cross-lingual analysis: accuracy and compactness of vowel categories \cite{kartushina2014effects, kartushina2016mutual, huffman2020relation, nycz2013best, dimov2013non}.
It has been frequently reported that L2 accent perception is highly correlated with these two factors \cite{huffman2020relation, kartushina2016mutual}. 
The vowel accuracy, or correctness, is the distance to the native productions of the target vowels \cite{dimov2013non}. 
% between the cross-lingually synthesized vowel and the native (non L2-accented) vowel on the vowel space.
Assuming that non cross-lingual TTS systems are not L2-accented, we measure the vowel accuracy by calculating the Euclidean distance between the non-accented target vowel and the corresponding vowel which is cross-lingually synthesized.
The greater vowel distance indicates the more L2-accented speech.
The vowel compactness, or variability, is one way to evaluate acoustic stability among vowel productions \cite{kartushina2016mutual}.
It is measured by calculating the standard deviation of each vowel.
The greater vowel standard deviation, or less compactness, implies the more L2-accented speech.
% Assuming that 
% We assume that 

% \section{Experiments}
% \label{sec:experiments}

% \begin{table}[t]
%   \caption{The mean vowel standard deviations across different systems and conditions.}
%   \label{tab:vowel_dev}
%   \centering
%   \begin{tabular}{c  c c c c }
%     \toprule
%     \multicolumn{1}{c}{\textbf{Model}} & 
%                                          \multicolumn{2}{c}{\textbf{Multi-speaker}} & \multicolumn{2}{c}{\textbf{Multi-lingual}} \\%&
%                                          %\multicolumn{1}{c}{\textbf{Set3}}\\
    
%     &shared&not shared&shared&not shared\\
%         \midrule

%     Tacotron     &0.426&0.439&0.356&0.492   \\
%     Glow-TTS     &0.325&0.346&0.285&0.298              \\

%     \bottomrule
%   \end{tabular}
  
% \end{table}

\section{Results and discussion}
\label{sec:results}

% \subsection{Probe \#1 by model configurations}

% \subsection{Probe \#2 model configuration}

% \subsection{Probe \#1 model configuration}

% The results from the vowel space analysis are shown in Table \ref{tab:vowel_distance}.

We first examine if L2 accents differ by the model architecture.
Table \ref{tab:vowel_distance} shows the mean vowel distance and the standard deviation of cross-lingually synthesized vowels from Tacotron and Glow-TTS.
Glow-TTS exhibits lower values in both metrics than Tacotron.
This implies that the L2 accent is more severe in Tacotron compared to Glow-TTS.
This phenomenon may arise from their different nature in auto-regressiveness, that is, auto-regressive systems are more influenced by preceding input tokens than its parallel counterpart during training.
However, further investigations are required to more clearly understand why Tacotron performs more poorly than Glow-TTS in cross-lingual speech synthesis. 
% Further investigations are required to .
% This may be because Glow-TTS is a newer model.
% Adding the language embedding shows a partial improvement in Tacotron, but does not show a clear improvement in Glow-TTS.
% This may be because of two reasons.
% First, cross-lingual speech synthesis attempts based on Tacotron have been more actively explored than in Glow-TTS.
% Second, the train database contains only one speaker for some languages. For these languages, the speaker/language information might not have been properly disentangled in the speaker/language embedding.
% We expect to 

\begin{table}[t]
\small
  \caption{The mean vowel distance and standard deviation across different TTS systems.
%   The smaller vowel distance and standard deviation indicate more accurate and stable vowel production, respectively.
  The shared vowels refer to the vowels that appear common in a language pair, while the non-shared vowels exist only in one language in a language pair.
  The smaller vowel distance and standard deviation indicate more accurate and stable vowel production, respectively.
%   The smaller values of both measurements in Glow-TTS and the shared vowels show that Tacotron and non shared vowels are more L2-accented.
}
  \label{tab:vowel_distance}
  \centering
  \begin{tabular}{c  c c c c}
    \toprule
    \multicolumn{1}{c}{\textbf{Model}} & \multicolumn{2}{c}{\textbf{Distance ($\downarrow$)}} &
                                         \multicolumn{2}{c}{\textbf{Standard Deviation ($\downarrow$)}} \\ %\multicolumn{2}{c}{\textbf{Multi-lingual}} \\%&
                                         %\multicolumn{1}{c}{\textbf{Set3}}\\
    
    &shared&non-shared&shared&non-shared\\
        \midrule

    % Tacotron     &0.796&0.920&0.720&0.824   \\
    % Glow-TTS     &0.571&0.671&0.584&0.715              \\
        Tacotron     &0.720&0.824&0.356&0.492   \\
    Glow-TTS     &0.584&0.715    &0.285&0.298          \\

    \bottomrule
  \end{tabular}
  
\end{table}

\begin{figure}[t]
    \centering

    % \begin{subfigure}[c]{.47\linewidth}
    %     \centering
    %     {\includegraphics[width=1.0\linewidth]{figures/dist_lpc_spk.pdf}}
    %     \caption{Multi-speaker Tacotron}\medskip
    %     \label{fig:into_a}
    % \end{subfigure}
    % \hfill
    \begin{subfigure}[c]{.47\linewidth}
        \centering
        {\includegraphics[width=1.0\linewidth]{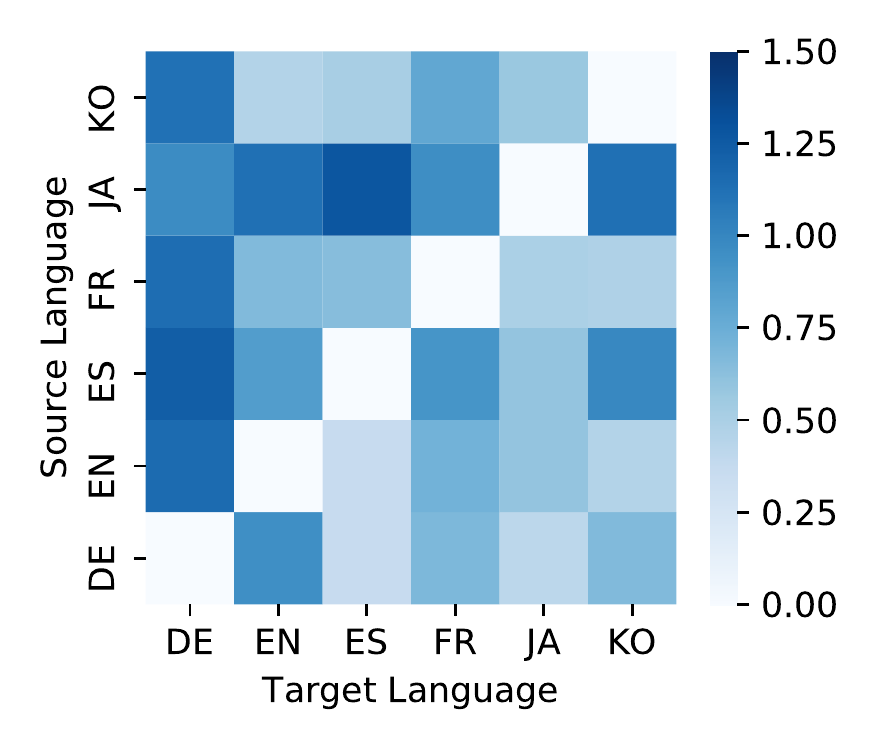}}
        \caption{Multi-lingual Tacotron}\medskip
        \label{fig:dist_lpc}
    \end{subfigure}
    \hfill
    % \newline
    % \begin{subfigure}[c]{.47\linewidth}
    %     \centering
    %     {\includegraphics[width=1.0\linewidth]{figures/dist_glow_spk.pdf}}
    %     \caption{Multi-speaker Glow-TTS}\medskip
    %     \label{fig:into_c}
    % \end{subfigure}
    % \hfill
    \begin{subfigure}[c]{.47\linewidth}
        \centering
        {\includegraphics[width=1.0\linewidth]{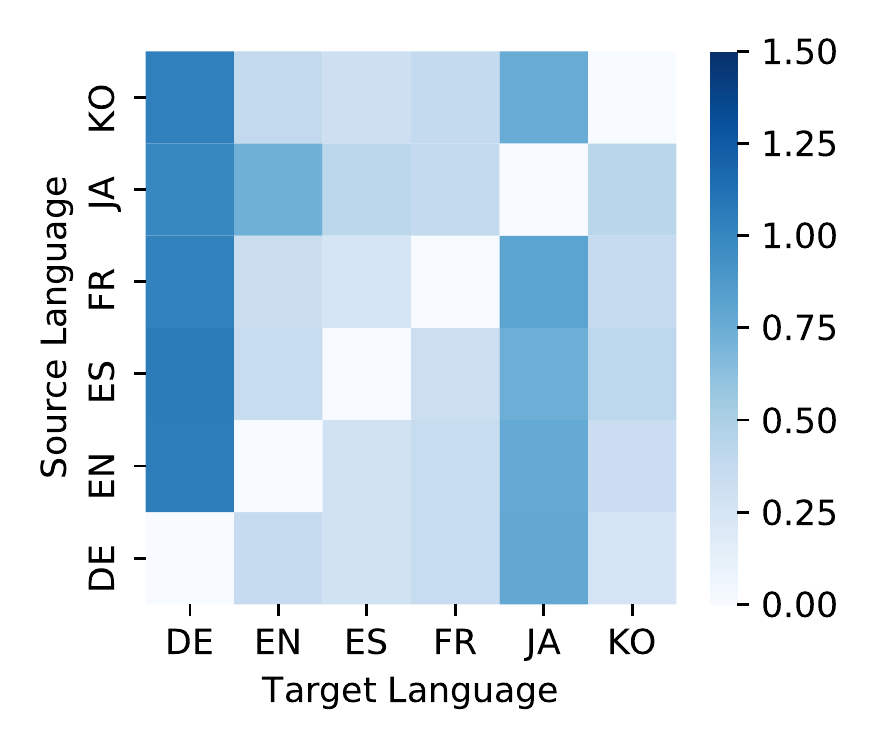}}
        \caption{Multi-lingual Glow-TTS}\medskip
        \label{fig:dist_glow}
    \end{subfigure}
%    \hfill
    \caption{The mean vowel distances in each language pair. The darker blue indicates the greater vowel distance, or the lesser vowel accuracy.
    Relatively low vowel accuracy is observed when the target language is German, or when the source language is Japanese or Spanish.
    }
    \label{fig:dist_voweldfdfd}
\end{figure}

We also explore if L2 accents of cross-lingual TTS systems are influenced by the linguistic characteristics among languages and if any of the observations appears in linguistics literature on actual human L2 learners.
We begin by comparing shared and non-shared vowels in a language pair.
The shared vowels are the vowels that appear in both languages in a language pair. For example, /i/ and /u/ are examples of the shared vowels in the English-German pair.
On the contrary, the non-shared vowels appear only in one language in a language pair, such as the German umlaut /\textipa{y}/.
As shown in Table \ref{tab:vowel_distance}, the vowels that are not shared in a language pair exhibit lower vowel accuracy and compactness, which indicate that the non-shared vowels are more prone to L2 accents.
The fact that unseen sounds are usually harder to process is one possible explanation for this result.
Considering that the shared vowels in different languages are regarded as different text input tokens, the chance is low that this result is caused by sharing the same text input tokens.
% is that unseen sounds are usually harder to process.
% This may be because totally unseen vowels are hard to synthesize, because no combination was trained.
A phenomenon similar to this is also observed in some studies on human L2 learners, that adult L2 learners exhibit difficulty with non-native phonological segments \cite{flege2003assessing}.

The vowel accuracy is the lowest when the target language is German regardless of the source language, as shown in Figure \ref{fig:dist_voweldfdfd}.
% This is may be because the German language has too many vowels.
It is also low when the source language is either Spanish or Japanese, especially in Tacotron.
This result may be related to the number of vowels each language has.
The German language has approximately 17 vowels (monophthongs), while Spanish or Japanese has only five.
When German is the target language, more unseen vowel sound categories should be created from the source language, and this is the most prominent in the languages with five vowels.
% This phenomenon is also observed in human L2 learners.
Similarly, human L2 learners experience more difficulty when the number of new sound categories they need to learn increases \cite{flege1997perpeption, weinberger1997minimal}.
% difficulty of learning a new language increases as the number of new sound categories they need to learn increases \cite{flege1997perpeption,weinberger1997minimal}.
Poor performance is observed when the target language is Japanese in Glow-TTS as in Figure \ref{fig:dist_glow}. This may arise from lack of explicit pitch accent information, which such parallel-based systems often require to properly model the Japanese language.

\begin{figure}[t]
    \centering
    \includegraphics[width=\linewidth]{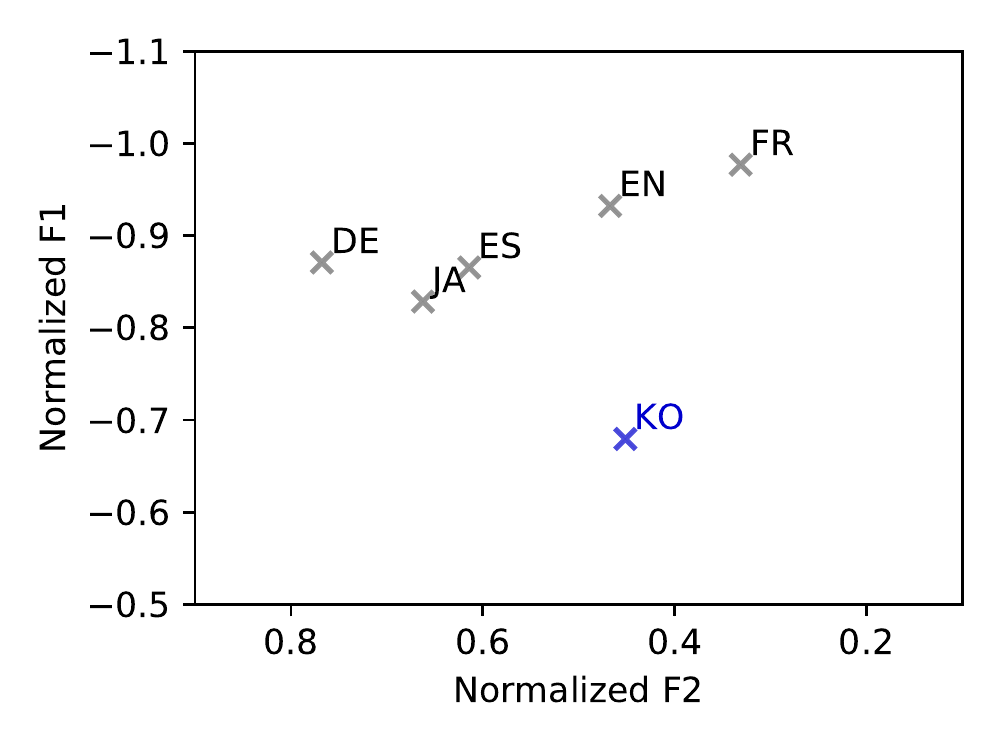}
    \caption{Cross-lingual realizations of the Korean vowel /\textipa{W}/ from various languages shown on the vowel space diagram.\protect\footnotemark}
    \label{fig:vowel_eu_kr}
\end{figure}

We inspect some vowels that are especially prone to L2 accents.
To list a few, the French nasal vowels and the Korean vowel /\textipa{W}/ rank top, probably because these vowels exist uniquely in each language.
Figure \ref{fig:vowel_eu_kr} depicts the Korean vowel /\textipa{W}/ on the vowel space along with other cross-lingually synthesized /\textipa{W}/ samples.
% Regardless of the source language, all of the non-native /\textipa{W}/ samples seem to have created a new vowel category, rather than merging into any neighboring vowels, as they are located reasonably distant from neighboring vowels such as /u/ or /i/, and this corresponds to \cite{flege1995second}.
All of the cross-lingually synthesized samples are located in the upper region compared to the native one, which means too much jaw closing.
The cross-lingual Korean /\textipa{W}/ when the source language is Japanese seems to be the least distant.
One possible reason is that the Japanese /u/ is usually realized as [\textipa{W}\textsuperscript{\textipa{B}}] (close back compressed vowel), slightly closer to the Korean /\textipa{W}/ \cite{vance2008sounds}.
Hence, when the source language is Japanese, the Korean /\textipa{W}/ is relatively less unseen compared to other European languages.
The realizations of /\textipa{W}/ tend to be more distant from that of Korean, when the source language is one of the three European languages (German, English, and Spanish), as they have distinct vowel systems from Korean.
In addition, some linguistics study suggests that Korean-English bilingual children speakers often assimilate the Korean /\textipa{W}/ to /u/ \cite{baker2005interaction}, and this phenomenon partly corresponds to our findings, as the jaw opening value ($F1$) of the cross-lingual /\textipa{W}/ from English is closer to /u/.

Our findings suggest that it may be necessary to consider linguistic characteristics among languages to mitigate the L2 accent issue.
For instance, an additional module, embedding, or loss function properly reflecting these linguistic characteristics in a language pair may help improve any L2 accents in cross-lingual scenarios.
These observations also imply we can utilize linguistics studies conducted on humans to develop TTS systems.
Our work may even be utilized as auxiliary information for building a new database or balancing an existing one.
% This is one of the most L2-accent-prone vowels, since it exists only in Korean.

% Lastly, we explore if any of our observations in the cross-lingual TTS systems appear in linguistics literature on actual human L2 learners.

% Now we answer the three research questions.

% \begin{figure}[t]
%     \centering

%     \begin{subfigure}[c]{.47\linewidth}
%         \centering
%         {\includegraphics[width=1.0\linewidth]{figures/dev_lpc_spk.pdf}}
%         \caption{Multi-speaker Tacotron}\medskip
%         \label{fig:into_a}
%     \end{subfigure}
%     \hfill
%     \begin{subfigure}[c]{.47\linewidth}
%         \centering
%         {\includegraphics[width=1.0\linewidth]{figures/dev_lpc_lang.pdf}}
%         \caption{Multi-lingual Tacotron}\medskip
%         \label{fig:into_b}
%     \end{subfigure}
% %    \hfill
%     \newline
    
%     \begin{subfigure}[c]{.47\linewidth}
%         \centering
%         {\includegraphics[width=1.0\linewidth]{figures/dev_glow_spk.pdf}}
%         \caption{Multi-speaker Glow-TTS}\medskip
%         \label{fig:into_c}
%     \end{subfigure}
%     \hfill
%     \begin{subfigure}[c]{.47\linewidth}
%         \centering
%         {\includegraphics[width=1.0\linewidth]{figures/dev_glow_lang.pdf}}
%         \caption{Multi-lingual Glow-TTS}\medskip
%         \label{fig:into_d}
%     \end{subfigure}
% %    \hfill
%     \caption{Mean Vowel Standard Deviation}
%     \label{fig:dev_all}
% \end{figure}

\footnotetext{Note that it is zoomed in for better visualization.}

\section{Conclusion and Future Work}
\label{sec:conclusion}

In this study, extending the vowel space analysis method to cross-lingual TTS systems, we explore L2 accents in such TTS systems.
We also compare some of observations on cross-lingual TTS systems with linguistics studies conducted on humans.
We hope our interdisciplinary study on cross-lingual TTS systems provides inspiration in relevant fields.

Based on what we have discovered, we are planning to mitigate the L2 accent problem in such cross-lingual TTS models, by proposing additional loss functions based on the linguistic characteristics in a language pair.

%\vfill
\clearpage
% \newpage
\pagebreak

% \vfill\pagebreak
% \pagebreak

% \section{REFERENCES}
% \label{sec:refs}

\ninept
\bibliographystyle{IEEEtrans}
\bibliography{refs}

\end{document}